# Minimum-sum dipolar spanning tree in $\mathbb{R}^3$

Steven Bitner and Ovidiu Daescu[*]
Department of Computer Science
University of Texas at Dallas
Richardson, TX 75080, USA

July 7, 2010

**Abstract**

In this paper we consider finding a geometric *minimum-sum dipolar spanning tree* in $\mathbb{R}^3$, and present an algorithm that takes $O(n^2 \log^2 n)$ time using $O(n^2)$ space, thus almost matching the best known results for the planar case. Our solution uses an interesting result related to the complexity of the common intersection of $n$ balls in $\mathbb{R}^3$, of possible different radii, that are all tangent to a given point $p$. The problem has applications in communication networks, when the goal is to minimize the distance between two hubs or servers as well as the distance from any node in the network to the closer of the two hubs.

## 1 Introduction

Let $S$ be a set of $n$ points in Euclidean Space. We study the *minimum-sum dipolar spanning tree* (MSST) problem, in which the goal is to find two points $x, y \in S$ that minimize the sum $|xy| + \max\{r_x, r_y\}$. This problem was first studied by Gudmundsson et al. [9].

Intuitively, the MSST is useful when one must choose two servers to service a set of clients, while the servers must also share data between them frequently. For example, it can be used in communication networks, when the goal is to minimize the distance between two hubs or servers as well as the distance from any node in the network to the closer of the two hubs, and could lead to reduction in power consumption for devices like PDAs, sensors, cell phones and laptops.

Gudmundsson et al. present exact results when $S$ is a set of $n$ points in $\mathbb{R}^d$, for $d \in \{2, 3, 4\}$ [9]. For the planar case, they show how to find the MSST in $O(n^2 \log n)$ time using $O(n^2)$ space. For dimensions $d = \{3, 4\}$, they suggest a solution based on range searching that takes $O(n^{2.5+\epsilon})$ time using $O(n^2)$ space, for any constant $\epsilon > 0$.

[*]Daescu's research was partially supported by NSF award CCF-0635013.

Let $G = (V, E)$ be a weighted undirected graph, where $V$ is the set of vertices, $E$ is the set of edges, and the weight function $W$ is defined on edges. The diameter of $G$ is the longest shortest path among all pairs of vertices of $G$. The *minimum diameter spanning tree* (MDST) problem asks to find a spanning tree $T$ of $G$ such that the maximum path length in $T$, over all pairs of vertices, is minimized, that is $\max\{\sum_{e \in p} W(e) | p \in T\}$ is minimized, where $p$ denotes a path between two vertices of $T$. A related problem, called the *Steiner* MDST, asks to connect a subset of nodes of $G$ while allowing additional vertices in the tree. That is, given $G = (V, E)$ and a subset $S$ of $G$, find a Steiner tree $T = (V_T, E_T)$ of $G$, with $S \subset V_T$, such that $\max\{\sum_{e \in p} W(e) | p \in T\}$ is minimized.

Minimum spanning tree and diameter problems on graphs have been studied for decades [3, 4, 6, 10, 12, 19] due to their various applications. For example, the MDST can be used to minimize the maximum distance required for a message to travel between any two nodes in a communication network, while keeping the number of network connections at a minimum.

In [12], Ho, Lee, Chang, and Wong introduce a special version of the MDST problem, where the vertices of $G$ are a set $n$ of points in the Euclidean space, and the edges of $G$ are implicitly defined by the pairwise distances between these points. The weight of an edge $e(u, v)$ connecting vertices $u$ and $v$ is defined as the Euclidean distance between the corresponding points. The resulting problem is called *geometric minimum diameter spanning tree* (GMDST). That is, for a set $S$ of $n$ points, the GMDST is defined as a spanning tree of $S$ that minimizes the Euclidean length of the longest path in the tree. They show there always exists a monopolar or a dipolar GMDST, i.e., a tree with only one or two vertices of degree greater than one, and give an $O(n^3)$ time algorithm for the planar case. Similarly, they define the *geometric Steiner minimum diameter spanning tree* (GSMDST) problem, in which Steiner points are allowed, and show that one Steiner point is sufficient for an optimal GSMDST. Thus, in the plane the problem is reduced to the minimum enclosing circle problem for a set of $n$ points, that can be solved in $O(n)$ time [7, 16].

It is interesting to notice that, even for the planar case, while a monopolar GMDST can be found in $O(n \log n)$ time [12] using the farthest point and second farthest point Voronoi diagrams [14, 15], finding a dipolar MDST efficiently has proven to be an elusive task [4, 12]. Specifically, for the dipolar MDST the goal is to find two points $x, y \in S$ that minimize the sum $r_x + |xy| + r_y$, where $|xy|$ is the Euclidean distance between the points $x$ and $y$, and $r_x$ and $r_y$ are the radii of two disks with centers at $x$ and $y$, respectively, that together cover all points in $S$. The best known result is based on semi-dynamic data structures and achieves $O^*(n^{3-c_d})$ time [4], where the $O^*$-notation hides an $O(n^\epsilon)$ term, for any constant $\epsilon > 0$, and $c_d = 1/((d+1)(\lceil d/2 \rceil + 1))$ is a constant that depends on the dimension $d$ of the point set. For example, $c_2 = 1/6$ and $c_3 = 1/12$.

A closely related problem is the *2-center* problem. In this problem, given a set $S$ of $n$ points in $\mathbb{R}^2$, the goal is to find two closed disks whose unions contain all points in $S$, and such that the radius of the larger disk is minimized. An $O(n^2 \log n)$ time result was given in [13] and was later improved by Sharir in [21] to a near linear time $O(n \log^9 n)$. Subsequently, using randomization, Eppstein

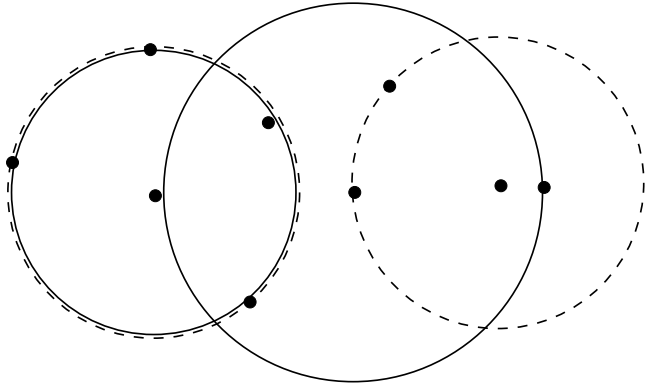

Figure 1: The solution for the discrete 2-center problem (dashed), and the MSST problem (solid).

achieved an expected time of $O(n \log^2 n)$ in [8].

The *discrete 2-center* problem where the center of the enclosing disks must lie on two points in $S$ was studied in [1], where the authors obtained an $O(n^{4/3} \log^5 n)$ running time.

## 1.1 Our Results

In this paper we consider finding the MSST in $\mathbb{R}^3$ and present an algorithm that takes $O(n^2 \log^2 n)$ time using $O(n^2)$ space, thus almost matching the best known results for the planar case. This problem looks to bridge the gap of the nearly cubic time of the *dipolar MDST* problem with the subquadratic time of the discrete 2-center problem. We note that during the execution of our algorithm, we can also obtain a solution to the discrete 2-center problem without any increase in the asymptotic running time. The converse of that is not true as can be seen in Figure 1.

We build on the algorithm given for the planar case in [9]. Our key contribution is an interesting result related to the complexity of the common intersection of balls that are all tangent to a given point. Specifically, given a set $S$ of $n$ balls in $\mathbb{R}^3$, of possible different radii, that are tangent to a given point $p$, we argue that their common intersection has complexity $O(n)$. This is in contrast to the $\Omega(n^2)$ complexity of the common intersection of $n$ balls of different radii, not restricted to be tangent to a common point.

## 1.2 Definitions and terminology

In this section we introduce some terminology and definitions. For two points $a$ and $b$, $|ab|$ denotes the Euclidean distance from $a$ to $b$. We use $\Sigma$ to denote a ball and $\Sigma(a, b)$ to denote the ball centered at $a$ and having $b$ on its bounding

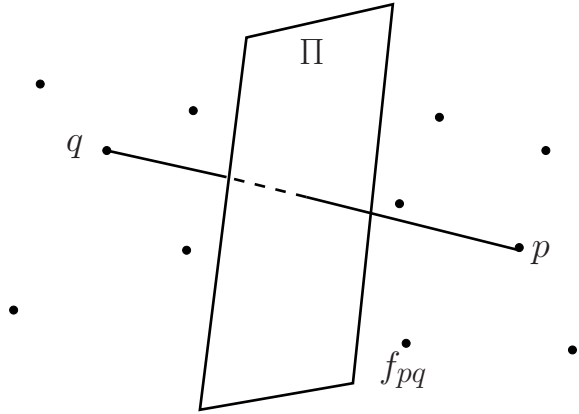

Figure 2: $p$, $q$, and the $q$-farthest point $f_{pq}$.

sphere, that is, the radius of $\Sigma(a,b)$ has length $|ab|$.

Let $p$ and $q$ be two points in $S$, and let $\Pi$ be the plane that is the perpendicular bisector of the line segment $\overline{pq}$. We use $h_{pq}$ to denote the open halfspace bounded by $\Pi$ and containing $p$. Similarly, $h_{qp}$ denotes the open halfspace bounded by $\Pi$ and containing $q$.

Given $p, q \in S$, the $q$-farthest point $f_{pq}$ is defined as the farthest point from $p$ that is contained in the open halfspace $h_{pq}$ (see Figure 2). A critical step in our solution is finding $f_{pq}$ for a fixed $p$ and all $q \in S \setminus \{p\}$ efficiently.

Given a sphere $\xi$ and a point $p$ on $\xi$, an inversion of $\xi$ with center of inversion $p$ maps $\xi$ into a plane by inverting the distance from $p$ to all points on the sphere [17]. Through inversion, using polar coordinates, a point $q = (\rho, \phi, \theta)$ is mapped to the point $q^* = (1/\rho, \phi, \theta)$, where $\rho$ is the distance from $p$ to $q$ and $\phi$ and $\theta$ are the polar angles. Notice that both polar angles are maintained through inversion. If $I$ denotes the inversion then $I(I(q)) = q$. Similarly, when applied to a plane that does not pass through $p$, the inversion yields a sphere which passes through $p$.

Consider the ball $\Sigma$ bounded by $\xi$. Let $\Pi$ be the plane corresponding to the inversion $I$. Then the interior of $\Sigma$ corresponds to one of the halfspaces bounded by $\Pi$.

## 2 Finding the MSST in $\mathbb{R}^3$

In this section we present our solution for finding a minimum-sum dipolar spanning tree in $\mathbb{R}^3$. To this end, we extend to $\mathbb{R}^3$ a lemma from [9] (Lemma 1 below) and give an algorithm to compute the MSST within the claimed time and space bounds. The algorithm makes use of a key property on the complexity of the common intersection of balls all tangent to a point $p$, that could be of interest beyond the scope of the MSST problem.

**Lemma 1.** *The point $x \in S$ is the $q$-farthest point from $p$ if and only if $x$ is the farthest point from $p$ satisfying $q \notin \Sigma(x, p)$.*

**Proof.** $\Rightarrow$ Since $x$ is the $q$-farthest point from $p$, by definition, it is contained in the open halfspace $h_{pq}$ and no other point of $S$ in $h_{pq}$ is farther from $p$ than $x$. Note that all points of $S \cap h_{qp}$ must have a smaller distance to $q$ than to $p$ since the halfspace $h_{qp}$ is defined by the orthogonal bisecting plane of $\overline{pq}$ (see Figure 2). That is, for any point $y \in S \cap h_{qp}$ the radius of the ball $\Sigma(y, p)$ is greater than $|yq|$, thus $q \in \Sigma(y, p)$. Then, $q \in \Sigma(x, p)$ would imply $|xq| < |xp|$, which means $x \in h_{qp}$, a contradiction. Thus, $q \notin \Sigma(x, p)$.

Since all points $x_i \in h_{qp}$ violate the statement that $q \notin \Sigma(x_i, p)$, and the $q$-farthest point is the point with the greatest distance from $p$ that is not in $h_{qp}$, the $q$-farthest point from $p$ is the farthest point from $p$ where $q \notin \Sigma(x, p)$.

$\Leftarrow$ Since $q \notin \Sigma(x, p)$, we have $|xp| < |xq|$. The half spaces $h_{pq}$ and $h_{qp}$ are defined by the perpendicular bisecting plane of $\overline{pq}$, so all points $y \in S$ with $|yp| < |yq|$ are contained in $h_{pq}$. (A similar argument can be made for those points in $h_{qp}$.) Thus, $x$ is the farthest point from $p$ among those in $S \cap h_{pq}$, which is precisely the definition for the $q$-farthest point. □

## 2.1 The Algorithm

Using Lemma 1 we can show that the approach presented in [9] for the planar case can be extended to $\mathbb{R}^3$. Specifically, for a fixed point $p \in S$, we can label all points $q \in S \setminus \{p\}$ with the $q$-farthest point $f_{pq}$ as follows.

First, sort $S$ in order of non-increasing distance from $p$. Second, set $f_{pq}$ for all points in $S$ to be NULL. Third, pass through the sorted array and for each point $q_i$, in order, set $f_{pq}$ to $q_i$ for all points $q \in S$ that are not contained by the ball $\Sigma(q_i, p)$ and for which $f_{pq}$ is set to NULL. That is, all points of $S$ that are in $\cap_{k=1}^{i-1} \Sigma(q_k, p)$ but not in $\Sigma(q_i, p)$, are labeled with $q_i$, where $i = 1, 2, \ldots, n-1$.

After the sorting above, the last value in the sorted array of points in $S \setminus \{p\}$ is the point which has minimum Euclidean distance from $p$. Therefore $D(q_n, p) \equiv D(p, p) = 0$. This implies that $f_{pq}$ is set for all points in $S \setminus \{p\}$. The sorted ordering also ensures that at any step in the algorithm, $f_{pq}$ for any point $q_i$ is the point corresponding to the smallest index $j$ for which $q_i \in \cap_{k=1}^{j-1} \Sigma(q_k, p)$ and $q_i \notin \Sigma(q_j, p)$. This means that given $q_i$ as $q$, $q_j$ is the farthest point from $p$ whose ball $\Sigma(q_j, p)$ does not contain $q_i$, which matches the definition of $f_{pq}$ for the given pair $p$, $q_i$.

It then follows that the generic algorithm for finding the $q$-farthest point for a fixed point $p$ and all $q \in S$ described in [9] for the planar case can also be applied in $\mathbb{R}^3$. We note that in $\mathbb{R}^3$ the problem of finding the smallest index $j$ for which $q_i \in \cap_{k=1}^{j-1} \Sigma(q_k, p)$ and $q_i \notin \Sigma(q_j, p)$ is also related to the *off-line ball exclusion testing* problem introduced in [2]. We present this algorithm below and then show how to perform the computations associated with it efficiently in $\mathbb{R}^3$, so that by applying it for each $p \in S$ we achieve the claimed time and space bounds.

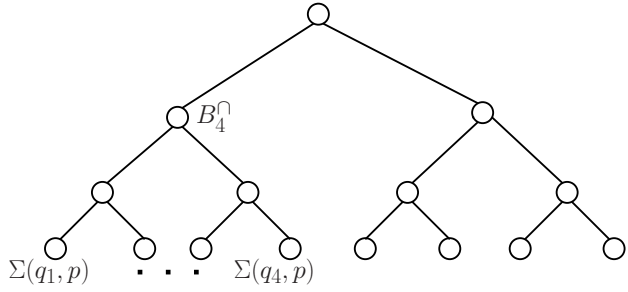

Figure 3: Binary search tree for $p$ used for finding $f_{pq}$ point for some input node $q$.

Without loss of generality, assume that $n = 2^k$ for some integer $k$. We build a complete binary tree $T$ with $k$ levels as follows. The leaves of $T$ are associated with the balls $\Sigma(q_i, p)$, $i = 1, 2, \ldots, n$, in order. That is, the leftmost leaf of $T$ stores $\Sigma(q_1, p)$ and the rightmost leaf of $T$ stores $\Sigma(q_n, p)$. Each internal node $v$ of $T$ stores a data structure associated with the common intersection of the balls that are leaf descendants of the sub-tree of $T$ rooted at $v$ (see Figure 3). Given a point $q$, to find the smallest index $j$ for which $q \in \cap_{k=1}^{j-1} \Sigma(q_k, p)$ and $q \notin \Sigma(q_j, p)$ start at the root of $T$ and follow a path to a leaf of $T$, at each node $v$ along the path performing the following test: if $q$ is in the common intersection stored at the left child of $v$ then go to the right child of $v$, else go to the left child of $v$. Clearly, the index associated with the leaf where this search ends corresponds to the sought $j$.

The common intersection of $n$ balls in $\mathbb{R}^3$, all having the same radius, has complexity $O(n)$ [11]. However, when the radii are not equal, which is our case, the common intersection can have complexity $\Omega(n^2)$. Thus, it is easy to check that a direct application of the algorithm above, with no other properties (like equal radii) in place, for each $p \in S$, would result in a solution for the MSST that takes cubic time and uses quadratic space, which is no better than brute force.

The astute reader may have noticed that answering whether a point $q$ is inside the common intersection of a set of balls in $\mathbb{R}^d$ may not require the actual computation of the common intersection of the balls. In fact, a ray shooting based approach to answer this query has been presented in [2], for solving a related problem termed off-line ball exclusion testing.

Barequet et al. [2] use a standard geometric mapping that lifts the point $q$ to a paraboloid in dimension $d+1$ and maps the balls into $(d+1)$-dimensional hyperplanes. The intersections of the hyperplanes with the paraboloid, projected back to dimension $d$, are the original balls. With this lifting, answering whether a point $q$ is inside the common intersection of $n$ balls in $\mathbb{R}^d$ is equivalent to answering whether a point in dimension $d+1$ is below the lower envelope of a set of $n$ $(d+1)$-dimensional hyperplanes. They showed that using a static data structure for ray shooting queries, that allows for trade-offs between the preprocessing time and the query time, answering the later question for a set of $n$ query points can be done in time and space $O(n^{2-2/(\lfloor (d+1)/2 \rfloor +1)} \log^{O(1)} n)$.

We can apply their solution on the nodes of $T$ resulting in an algorithm that takes $O(n^{2-2/(\lfloor (d+1)/2 \rfloor +1)} \log^{O(1)} n)$ time and space. Since we have to do this once for each $p \in S$, the time to find the MSST is $O(n^{3-2/(\lfloor (d+1)/2 \rfloor +1)} \log^{O(1)} n)$. Each ray shooting data structure can be discarded after serving its purpose, so the overall space requirement remains $O(n^2)$, due to storing $f_{pq}$ for each pair of points $p, q \in S$.

Thus, for any constant dimension $d$, we obtain the following result.

**Lemma 2.** *Given a set $S$ of $n$ points in $\mathbb{R}^d$, $d \geq 3$ a constant, the MSST of $S$ can be found in $O(n^{3-2/(\lfloor (d+1)/2 \rfloor +1)} \log^{O(1)} n)$ time and $O(n^2)$ space.*

For $d = 3, 4$, this gives an algorithm for the MSST with running time of $O(n^{7/3} \log^{O(1)} n)$ which improves over the $O(n^{2.5+\epsilon})$ time algorithm in [9].

We will show in the next subsection that a faster solution can be obtained in $\mathbb{R}^3$ by actually computing the common intersection of the balls stored at internal nodes of $T$.

## 2.2 Intersection of Balls Tangent to a Point

Consider the common intersection of a set $B$ of $n$ balls $\Sigma_1, \Sigma_2, \ldots, \Sigma_n$ in $\mathbb{R}^3$, all tangent to a point $p$. We have the following property.

**Lemma 3.** *Each ball in $B$ can contribute at most one connected component to the boundary of the common intersection.*

**Proof.** Let $a$ and $b$ be two points on the boundary of the common intersection $bd(B^\cap)$ of the balls in $B$, both on the same bounding sphere $s$ of some ball in $B$. The plane defined by $a$, $b$, and $p$ intersects $s$ in a circle $c$. The geodesic connecting $a$ and $b$ along $c$ on $s$ (the arc $\widehat{ab}$ of $c$) must be in $bd(B^\cap)$; otherwise, if another ball contains $a$ and $b$ but not some other point $q$ on $\widehat{ab}$, then the bounding sphere $s'$ of that ball defines a circle $c'$ in the plane of $a$, $b$, and $p$ that has radius greater than that of $c$ and contains $p$ (see Fig. 4), a contradiction to the fact that $s'$ is tangent to $p$. Thus, $bd(B^\cap) \cap s$ has at most one connected component. □

Assuming general position (that is, no more than three bounding spheres intersect in a point other than $p$), Lemma 3 implies the complexity of $bd(B^\cap)$ is $O(n)$. We notice that a similar result can be derived from [17] using inversion.

Let $T$ be a tree as described in Section 2.1. The intersection of the balls associated with the internal nodes of $T$ can be computed in a bottom-up fashion, using the algorithm in [20]. Although that algorithm was designed for equal radius balls, we note that the only place in that algorithm where equal radii plays a role is in obtaining the property that each ball contributes only one connected component to $bd(B^\cap)$. The algorithm computes the common intersection at each internal node by *merging* the intersections stored at its children and takes $O(n \log^2 n)$ time over $T$.

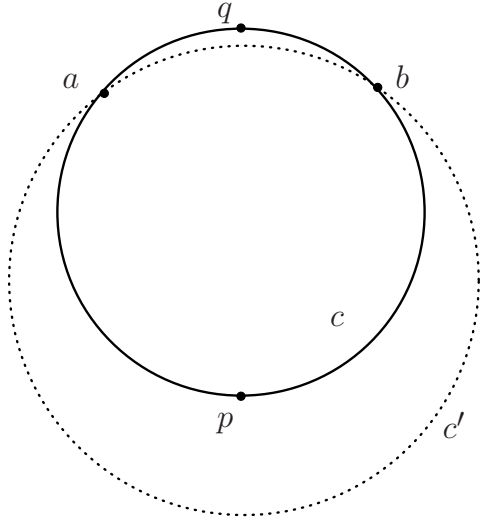

Figure 4: Circles $c$ and $c'$ in the plane of $a, b, p$.

However, we can do better by using inversion. To this end, let $T$ be a tree as above. Instead of storing the balls at the leaves of $T$, we store the corresponding halfspaces, obtained by using $p$ as the center of inversion.

**Lemma 4.** *The tree $T$ can be constructed in $O(n \log n)$ time and uses $O(n \log n)$ space.*

**Proof.** Invert all balls using $p$ as the center of inversion and store the resulting halfspaces at the corresponding leaf nodes. This takes $O(n)$ time and space. The balls in the original problem become halfspaces in the inversion space. The intersection of $n$ balls corresponds to the convex polytope that is the intersection of the halfspaces. The complexity of this polytope is $O(n)$. The faces of the polytope can be inverted back in $O(n)$ time to obtain the spherical portions of the intersection of the balls in the problem space. (Notice that since each spherical portion in the common intersection of the balls corresponds to a face in the inversion space, and the complexity of all faces is $O(n)$, it follows that the intersection of the balls also has complexity $O(n)$.)

Let $v$ be an internal node of $T$. We can obtain the polytope associated with $v$ by computing the common intersection of the polytopes associated with the left and right children of $v$, which takes linear time and space in the complexity of the children [5].

We can store both the polytope and the ball intersection at $v$ without asymptotically increasing the space requirements. However, for our purpose, we only need to store the polytope. By performing a bottom-up traversal of $T$, the overall time to compute the common intersections (the polytopes) for all nodes in $T$ is thus $O(n\log n)$. Similarly, the space requirement is $O(n\log n)$. □

**Lemma 5.** *Given a set of n balls all tangent to a point p, and a query point q, the smallest index i such that $\Sigma_i$ does not contain q can be found in $O(\log^2 n)$ time.*

**Proof.** Let $q^*$ be the inversion of $q$ with center $p$. We use the complete binary tree $T$ described earlier enhanced with point location capability at each internal node.

Specifically, when traversing $T$ on a path from the root to a leaf, we need to decide at each internal node $v$ on the path whether $q^*$ is inside the convex polytope associated with the left child of $v$. If it is, we descend to the right child, otherwise we descend to the left child.

Assume the left child $u$ of $v$ has $m$ leaf descendants. Having stored at $u$ a point location data structure that requires $O(m)$ space and can be constructed in $O(m)$ time, the query at $v$ can be answered in $O(\log m)$ time [18] (page 285). Over all nodes in $T$ the point location data structures can be built in $O(n \log n)$ time using $O(n \log n)$ space, which is done as a preprocessing step.

Thus, the overall query time along the root-to-leaf path is $O(\log^2 n)$. □

Since the data structure for $p$ can be discarded after $f_{pq}$ is found for each $q \in S \setminus \{p\}$, we obtain:

**Theorem 1.** *Given a set S of n points in $\mathbb{R}^3$, the MSST of S can be found in $O(n^2 \log^2 n)$ time using $O(n^2)$ space.*

We also mention that from an implementation viewpoint this solution should be easier to implement than the ray shooting based solution. Moreover, all of the geometric primitives required by this algorithm have been implemented in various geometric packages and are readily available for use.

## 3 Conclusion

In this paper we presented an $O(n^2 \log^2 n)$ time, $O(n^2)$ space algorithm for finding a geometric minimum-sum dipolar spanning tree in $\mathbb{R}^3$, almost matching the $O(n^2 \log n)$ time for the planar case. In the process, we argued that the common intersection of $n$ balls in $\mathbb{R}^3$, of possible different radii, that are all tangent to a point $p$, has complexity $O(n)$. This is in contrast to the $\Omega(n^2)$ complexity of the common intersection of $n$ balls of different radii, not restricted to be tangent to a common point.

The $n$x$n$ matrix that we compute during this algorithm can easily be used to return a solution to the discrete 2-center problem in quadratic time. Thus, we implicitly provide a solution to the discrete 2-center problem in $\mathbb{R}^3$. This solution runs in $O(n^2 \log^2 n)$ time and uses quadratic space.

We notice that the extra $\log n$ time in $\mathbb{R}^3$, when compared to the planar counterpart, comes from the query phase of the algorithm. If the time for querying the tree $T$ with a point $q$ can be reduced to $O(\log n)$ then the algorithm would match the time and space complexities of the planar version.

# References

[1] P. Agarwal, M. Sharir, and E. Welzl. The discrete 2-center problem. *Discrete & Computational Geometry*, 20(3):287–305, 1998.

[2] G. Barequet, D.Z. Chen, O. Daescu, M.T. Goodrich, and J. Snoeyink. Efficiently approximating polygonal paths in three and higher dimensions. *Algorithmica*, 33(2):150–167, 2002.

[3] W.G. Brown, editor. *Reviews in Graph Theory*. American Mathematical Society, Providence, RI, USA, 1980.

[4] T.M. Chan. Semi-online maintenance of geometric optima and measures. *SIAM Journal on Computing*, 32(3):700–716, 2003.

[5] B. Chazelle. An optimal algorithm for intersecting three-dimensional convex polyhedra. *SIAM Journal on Computing*, 21(4):671–696, 1992.

[6] D. Cheriton and R.E. Tarjan. Finding minimum spanning trees. *SIAM Journal on Computing*, 5:724–742, 1976.

[7] M. Dyer. Linear time algorithms for two- and three-variable linear programs. *SIAM Journal on Computing*, 13(1):31–45, 1984.

[8] D. Eppstein. Faster construction of planar two-centers. In *SODA '97: Proceedings of the eighth annual ACM-SIAM symposium on Discrete algorithms*, pages 131–138, Philadelphia, PA, USA, 1997. Society for Industrial and Applied Mathematics.

[9] J. Gudmundsson, H. Haverkort, S.-M. Park, C.-S. Shin, and A. Wolff. Facility location and the geometric minimum-diameter spanning tree. *Lecture Notes in Computer Science*, 2462:146–160, 2002.

[10] J. Gudmundsson, H. Haverkort, S.-M. Park, C.-S. Shin, and A. Wolff. Facility location and the geometric minimum-diameter spanning tree. *Computational Geometry: Theory and Applications*, 27(1):87–106, 2004.

[11] A. Hepes. Beweis einer Vermutung von A. Vazsonyi. *Acta Mathematica Academiae Scientiarum Hungaricae*, 7:463–466, 1956.

[12] J.-M. Ho, D.T. Lee, C.-H. Chang, and C.K. Wong. Minimum diameter spanning trees and related problems. *SIAM Journal on Computing*, 20(5):987–997, 1991.

[13] J. Jaromczyk and M. Kowaluk. An efficient algorithm for the euclidean two-center problem. In *Proceedings of the tenth annual symposium on Computational geometry*, pages 303–311, New York, NY, USA, 1994. ACM.

[14] D.T. Lee. Farthest neighbor voronoi diagrams and applications. Technical Report 80-11-FC-04, Northwestern University, Evanston, IL, 1980.

[15] D.T. Lee. On finding $k$-nearest neighbor voronoi diagrams in the plane. Technical report, 1982.

[16] N. Megiddo. Linear time algorithms for linear programming in $\mathbb{R}^3$ and related problems. *SIAM Journal on Computing*, 12:759–776, 1983.

[17] K. Mehlhorn. *Data structures and algorithms 3: multi-dimensional searching and computational geometry*. Springer-Verlag, New York, NY, USA, 1984.

[18] J. O'Rourke. *Computational Geometry in C, Second Edition*. Cambridge University Press, New York, NY, USA, 1998.

[19] R. Prim. Shortest connecting networks and some generalizations. *Bell System Technical Journal*, 36:1389–1401, 1957.

[20] E.A. Ramos. Intersection of unit-balls and diameter of a point set in $\mathbb{R}^3$. *Computational Geometry*, 8:57–65, 1997.

[21] M. Sharir. A near-linear algorithm for the planar 2-center problem. *Discrete & Computational Geometry*, 18(2):125–134, 1997.